# AlH$_3$ between 65 -110 GPa: implications of electronic band and phonon structures

A.K.M.A. Islam[1], M.M. Ali, and M.L. Ali

Department of Physics, Rajshahi University, Rajshahi 6205, Bangladesh

**Abstract**

A first-principles density-functional-theory method has been used to reinvestigate the mechanical and dynamical stability of the metallic phase of AlH$_3$ between 65-110 GPa. The electronic properties and phonon dynamics as a function of pressure are also explored. We find electron-phonon superconductivity in the cubic *Pm*-3*n* structure with critical temperature $T_c$ = 37 K at 70 GPa which decreases rapidly with the increase of pressure. Further unlike a previously calculated $T_c$ value of 24 K at 110 GPa, we do not find any superconductivity of significance at this pressure which is consistent with experimental observation.

**Keywords**: AlH$_3$, Electronic structure, Phonon dynamics, Superconductivity

PACS number(s):  62.50.-p, 63.20.kd, 74.10.+v, 74.20.Pq

## 1. Introduction

The hydrogen-rich (10% by weight) solid alane (AlH$_3$) has recently been the focus of renewed interest among the scientific community [1-15]. The crystal structure of the ambient phase has been studied by Turley and Rinn [1]. The most stable structure of AlH$_3$ (space group *R*-3*c)* has been found to be stable for 0-35 GPa [3]. Scheicher *et al.* [11] found a cubic phase of space group *Pm*-3*n* which becomes energetically favourable at above 60 GPa and it exhibits metallic behaviour. On the other hand Pickard and Needs [8] found the same phase to be favourable for $P \geq 73$ GPa. According to them the semimetallic phase at 73 GPa becomes metallic above 80 GPa. Vajeeston et al. [14] have shown that application of pressure makes sequence of phase transitions from $\beta \rightarrow \alpha' \rightarrow \alpha \rightarrow hp1$ (*P*6$_3$/*m* ) $\rightarrow$ *hp*2 (*Pm*-3*n*) modification at pressures of  2.4, 4.3, 64, and 104 GPa, respectively. The *hp*1 phase has a hexagonal (*P*6$_3$/*m*) structure with the lowest energy among all these phases and the orthorhombic (*Pnma*) structure suggested in [8] is energetically closer to the *P*6$_3$/*m* structure. The electronic structures reveal that $\alpha, \alpha', \beta$ polymorphs are non-metals, whereas the *P*6$_3$/*m* and *Pm*-3*n* phases possess semiconducting and metallic behaviour, respectively. Goncharenko *et al.* [12] found theoretically that AlH$_3$ transforms to a lower symmetry and lower enthalpy triclinic structure formed by distorted and shifted triangular Al planes at pressures above 60 GPa. They found that a simple cubic structure becomes energetically favourable at pressures above 100 GPa.  The structure *hp*1 (*P*6$_3$/*m*) phase has not yet been solved experimentally, as it can be either monoclinic or triclinic [12].

Thus it is obvious that there are some differences in the pressure-dependent structures and structural transitions predicted by different studies [8, 11-14]. Further the theoretical prediction about the sign of a superconductive transition at ~ 110 GPa contrasts with the experimental observation [12]. All these make the system an interesting case for further analysis. The purpose of the present DFT calculations is to shed further light on the behaviour of the high-pressure phase of AlH$_3$ particularly between 65-110 GPa and to focus more on the situation regarding superconductivity based on the electronic properties and phonon dynamics.

## 2. Computational Methods

Our calculations were performed using the *ab initio* plane-wave pseudopotential approach within the framework of the density-functional theory implemented in the CASTEP software [16]. The ultrasoft pseudopotentials were used in the calculations, and the plane-wave cutoff energy was 400 eV. The exchange-correlation terms used are of the Perdew-Berke-Ernzerhof form of the generalized gradient approximation [17]. The $k$-points samplings were 12×12×12 in the Brillouin zone for the cubic $AlH_3$, according to the Monkhorst-Pack scheme. All the structures were relaxed by the BFGS methods [18] where the convergence criterion used is $0.5×10^{-5}$ eV/atom. The elastic constants $C_{ij}$, bulk modulus $B$ and electronic properties were directly calculated by the CASTEP code.

---

[1]Corresponding author.
*E-mail address*: azi46@ru.ac.bd (A.K.M.A. Islam). *Fax:* +88 0721 750064, *Tel*: +88 0721 750980

Calculations of phonon spectra, electron-phonon (e-ph) coupling and phonon density of states were performed using plane waves and pseudopotentials with Quantum Espresso [19]. We employed ultrasoft Vanderbilt pseudopotentials [20], with a cutoff of 70 Ryd for the wave functions, and 700 Ryd for the charge densities. The $k$-space integration for the electrons was approximated by a summation over a 12×12×12 uniform grid in reciprocal space, with a Gaussian smearing of 0.03 Ryd for self-consistent cycles. Dynamical matrices and e-ph linewidths were calculated on a uniform 3×3×3 grid in phonon $q$-space. Phonon dispersions and DOS were then obtained by Fourier interpolation of the dynamical matrices

## 3. Results and Discussions

### 3.1 Stability of high-pressure phase

The prevalent ambient phase is the hexagonal phase $\alpha$-$AlH_3$ which crystallizes in the space group $R$-$3c$. The lattice constants found from full relaxation at ambient pressure are $a = b = 4.4991$ Å, $c = 11.8617$ Å, and $\gamma = 120°$, which are in good agreement with measured values [1] and theoretical calculation [14]. In view of our previous comments the intention is to check first whether other structures compete with the cubic $AlH_3$ for pressure above 60 GPa. So we consider the following most probable structures so far found [1, 8, 9, 12, 13, 15] at and above zero pressures: $\alpha$-phase, $\gamma$-phase, $Pnma$, $Pm$-$3n$, $Pbcm$, $P63/m$. The calculated enthalpies with respect to the ambient hexagonal phase $\alpha$-$AlH_3$ (space group $R$-$3c$) as a function of pressure are shown in Fig. 1(a). On comparison, it is found that a cubic phase of space group $Pm$-$3n$ (Fig. 1 (b)) becomes energetically favourable for $P \geq 69$ GPa. This value of 69 GPa can be compared with values of 73 GPa [8] and 72 GPa [13]. Such kind of variation may arise due to use of different codes as indicated by Kim *et al*. [13] with ABINIT and VASP. But the variation between our value and that in [8] is due to the use of different optimization parameters (for example among others we used a higher plane wave cutoff energy of 400 eV compared to 260 eV in [8]).

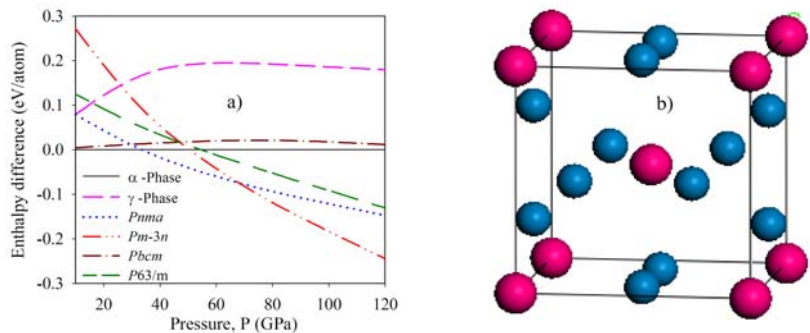

Fig. 1. (a) Enthalpy difference *versus* pressure for six low enthalpy phases of $AlH_3$, and (b) unit cell of cubic $AlH_3$ in space group $Pm$-$3n$ (Al at 2a sites with larger pink-coloured spheres; H at 6c sites with smaller blue-coloured spheres).

For a crystal with cubic symmetry, the mechanical stability is that the tetragonal and trigonal shear elastic constants $C_- = (C_{11}-C_{12})/2$ and $C_{44}$ are both positive [21]. Using our calculated elastic constants we find that both of these quantities increase proportionally to the applied pressure, and remain positive throughout the investigated pressure range 65 -110 GPa. Thus the cubic phase of $AlH_3$ is mechanically stable regardless of the applied pressure.

### 3.2 Electronic band structure

The band structures of cubic $AlH_3$ at three different pressures are shown in Fig. 2 (a, b, c). The calculations yield indirect band gap overlap of 1.56, 1.37, 1.31 eV at 70, 100 and 110 GPa, respectively. The Fermi surface of *Pm*-3*n* contains electronic states of very different character. As observed earlier [8] we find the band minimum near the Fermi level at the *R* point to be highly dispersive and the state is associated with the Al atoms. At the *M* point the band maximum is associated with the chains of H atoms. This shows substantial energy overlap with the Al-derived band at *R*. The results of the calculated total electronic DOS are displayed in Fig. 2 (d). The inset shows the magnified DOS values at the Fermi level. That $AlH_3$ exhibits metallic character from 70 to 110 GPa can be seen from the band structures and the finite DOS values at the Fermi energy. The values of DOS (states/eV) at 70, 100, 110 GPa are 0.502, 0.370 and 0.358, respectively.

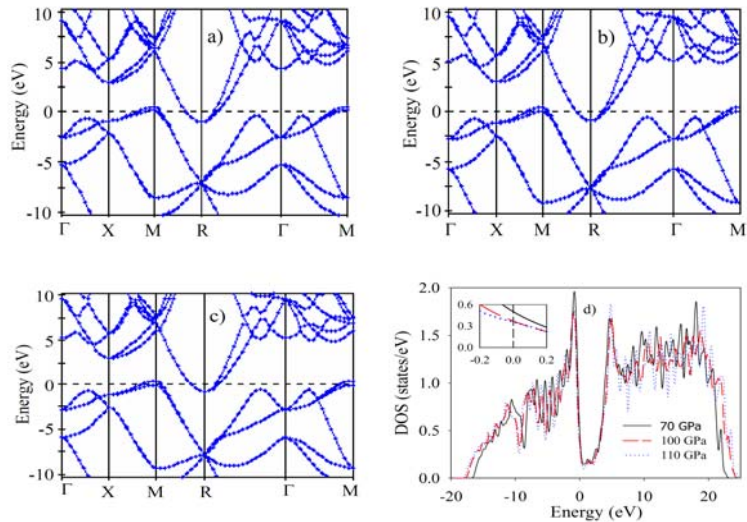

Fig. 2. Band structure of cubic $AlH_3$ at (a) 70 GPa, (b) 100 GPa, (c) 110 GPa; (d) Total DOS at three different pressures. Inset: expanded view at Fermi level.

### 3.3 Phonon spectrum and electron-phonon coupling

We have calculated the phonon dispersions of the cubic phase $AlH_3$ for $69 \geq P \leq 100$ GPa using Quantum Espresso code. This would help us to analyse the dynamical stability and to compare our results with those of Kim *et al.* [13] who used ABINIT code. The phonon-dispersion relation for two pressure values (70 and 100 GPa) are shown in Fig. 3. The phonon modes are seen to be grouped into two parts: the lower bands, range up to ~500 cm$^{-1}$ and ~550 cm$^{-1}$, for 70 and 100 GPa, respectively. The lower band is mostly attributed due to Al while the higher band is formed by Al-H and H-H interactions. The lowest *X* mode is seen to become softer at lower pressure.

The corresponding phonon density of states is shown on the right panel of each of the figures for 70 and 100 GPa. It is clear that all the phonon frequencies are real and no imaginary phonon frequency is observed in the whole Brillouin zone in the calculated phonon dispersion curves and phonon density of states of $AlH_3$ at 70 and 100 GPa. The same is true for intermediate pressure 80

GPa and also at pressure 110 GPa (figures not shown). This indicates that AlH$_3$ structure is dynamically stable at these pressures. An analysis of the lattice dynamics by Kim *et al.* [13] showed that the system is dynamically stable in the pressure range of 72–106 GPa.

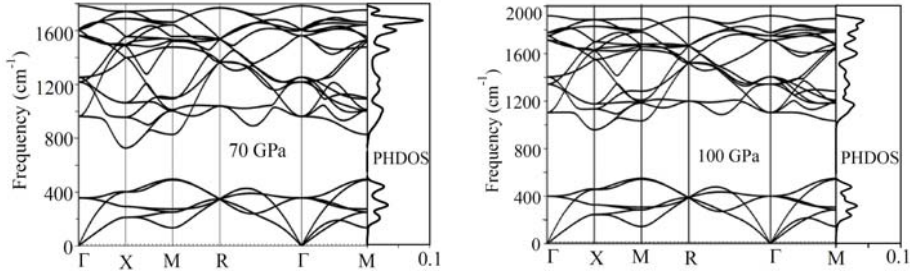

Fig. 3. Phonon dispersion and projected phonon DOS of cubic AlH$_3$ at (a) 70 GPa, (b) 100 GPa.

The frequency-dependent electron-phonon coupling constant $\lambda(\omega)$ and Eliashberg function $\alpha^2F(\omega)$ are given by [22]

$$\lambda(\omega) = 2 \int_0^\omega d\Omega\, \alpha^2 F(\Omega)/\Omega \tag{1}$$

$$\alpha^2 F(\omega) = \frac{1}{N(0)} \sum_{nm\mathbf{k}} \delta(\varepsilon_{n\mathbf{k}}) \delta(\varepsilon_{m\mathbf{k}+\mathbf{q}}) \sum_{\nu \mathbf{q}} \left| g_{\nu,n\mathbf{k},m(\mathbf{k}+\mathbf{q})} \right|^2 \delta(\omega - \omega_{\nu\mathbf{q}}) \tag{2}$$

where $g_{\nu,n\mathbf{k},m\mathbf{k}'}$ are the bare DFT matrix elements.

The total e-ph coupling constant $\lambda$ for metallic state of AlH$_3$, is obtained by numerical integration of equation (1) up to $\omega = \infty$, which yields $\lambda$ and the corresponding values for the logarithmically-averaged frequencies ($\omega_{ln}$). We use the Coulomb pseudopotential $\mu^* = 0.14$ and the simplified Allen-Dynes formula for $T_c$ [23] within the phonon mediated theory of superconductivity:

$$k_B T_c = \frac{\hbar \omega_{ln}}{1.2} \exp\left[ -\frac{1.04(1+\lambda)}{\lambda - \mu^*(1+0.62\lambda)} \right] \tag{3}$$

The results of calculations are shown in Fig. 4. We see that the finite value of $T_c = 37$ K at 70 GPa goes down quickly with the increase of pressure. It is worth mentioning that the calculated DOS at the

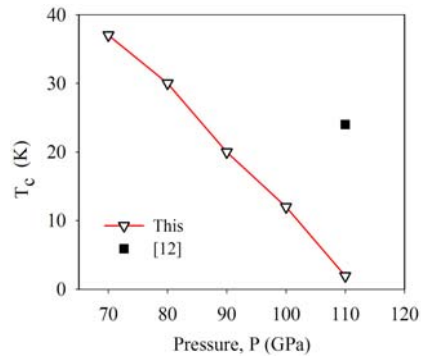

Fig. 4. Calculated $T_c$ as a function of pressure for cubic AlH$_3$.

Fermi level decreases with pressure. We may also add here that the calculated softening of $X$ point phonon modes decreases with the increase of pressure (see Fig. 3) due to Fermi surface nesting [12, 24]. All these would imply that $T_c$ of AlH$_3$ would decrease with pressure. This trend is in agreement with our calculated $T_c$-values. Further at 110 GPa the $T_c$-value becomes almost zero. It is to be noted here that Goncharenko et al. [12] employed the above equation with $\mu^* = 0.14$ to predict the critical temperature ($T_c$) as 24 K at 110 GPa. Thus it is clear that unlike the prediction of these authors, we do not find any superconductivity of significance which is consistent with experimental observation [12].

We recall here that Pickard and Needs [8] performed DFT calculations of phonon frequencies of the Pm-3n structure at 80 GPa. They then obtained a Debye temperature $\theta_D$ as 2700 K by assuming the highest calculated phonon frequency as a measure of the Debye frequency. They have commented, without calculating $T_c$, that the high value of $\theta_D$ would favour superconductivity, but the possibility diminishes due to rather small total density of electronic states at the Fermi energy.

## 4. Conclusion

The metallicity, mechanical and dynamical stability of AlH$_3$ phase between 65-110 GPa have been reanalysed in view of some differences in the pressure-dependent structures and structural transitions predicted by different studies. We focus more on the situation regarding superconductivity and find electron-phonon superconductivity in the cubic Pm-3n structure with critical temperature of 37 K at 70 GPa. The value decreases rapidly with the increase of pressure. Unlike a previously calculated $T_c$ value of 24 K at 110 GPa, we do not find any superconductivity of significance at this pressure which is consistent with experimental observation. The decreasing $T_c$-value at higher pressure may be due to the decrease of softening of $X$ point phonon modes as a result of Fermi surface nesting.

**Acknowledgements**

The authors acknowledge Rajshahi University and the Bangladesh University Grants Commission for providing facilities to carry out the work.

**References**

[1]   J. W. Turley, H. W. Rinn, Inorg. Chem. 8 (1969) 18-22.
[2]   F. M. Brower, N. E. Matzek, P. F. Reigler, H.W. Rinn, C. B. Roberts, D. L. Schmidt, J. A. Snover, and K. Terada, J. Am. Chem. Soc. 98 (1976) 2450-2453.
[3]   B. Baranowski, H. D. Hochheimer, K. Strossner, and W. Honle, J. Less-Common Met. 113, 341 (1985).
[4]   X. Ke, A. Kuwabara, and I. Tanaka, Phys. Rev. B 71 (2005) 184107 [7 pages].
[5]   J. Graetz, S. Chaudhuri, Y. Lee, and T. Vogt, Phys. Rev. B 74 (2006) 214114 [7 pages].
[6]   H.W. Brinks, A. Istad-Lem, and B. C. Hauback, J. Phys. Chem. B 110 (2006) 25833-25837.
[7]   H.W. Brinks, A. Istad-Lem, and B. C. Hauback, J. Alloys Compd. 433 (2007) 180-183.
[8]   C. J. Pickard, and R. J. Needs, Phys. Rev. B 76 (2007) 144114 [5 pages].
[9]   V. A. Yartys, R.V. Denys, J. P. Maehlen, C. Frommen, M. Fichtner, B.M. Bulychev, and H. Emerich, Inorg. Chem. 46 (2007) 1051-1055.
[10]  H.W. Brinks, C. Brown, C. M. Jensen, J. Graetz, J. J. Reilly, and B. C. Hauback, J. Alloys Compd. 441 (2007) 364-367.
[11]  R. H. Scheicher, D. Y. Kim, S. Lebègue, B. Arnaud, M. Alouani, and R. Ahuja, App. Phys. Letters 92 (2008) 201903 [3 pages].
[12]  I. Goncharenko, M. I. Eremets, M. Hanfland, J. S. Tse, M. Amboage, Y. Yao, and I. A. Trojan, Phys. Rev. Letters 100 (2008) 045504 [4 pages].
[13]  D. Y. Kim, R. H. Scheicher, and R. Ahuja, Phys. Rev. B 78 (2008) 100102(R) [4 pages].
[14]  P. Vajeeston, P. Ravindran, and H. Fjellvåg, Chem. Mater., 20 (19) (2008) 5997-6002.
[15]  S. Sun, X. Ke, C. Chen, and I. Tanaka Phys. Rev. B 79 (2009) 024104 [8 pages].
[16]  S. J. Clark, M. D. Segall, C. J. Pickard, P. J. Hasnip, M. J. Probert, K. Refson, and M. C. Payne, Zeitschrift fuer Kristallographie 220 (2005) 567-570.
[17]  J. P. Perdew, K. Burke, and M. Ernzerhof, Phys. Rev. Lett. 77 (1996) 3865-3868.
[18]  T. H. Fischer and J. Almlof, J. Phys. Chem. 96 (1992) 9768-9774.

[19] P. Giannozzi, S. Baroni, N. Bonini, M. Calandra, R. Car, C. Cavazzoni, D. Ceresoli, G. L. Chiarotti, M. Cococcioni, I. Dabo, A. Dal Corso, S. Fabris, G. Fratesi, S. de Gironcoli, R. Gebauer, U. Gerstmann, C. Gougoussis, A. Kokalj, M. Lazzeri, L. Martin-Samos, N. Marzari, F. Mauri, R. Mazzarello, S. Paolini, A. Pasquarello, L. Paulatto, C. Sbraccia, S. Scandolo, G. Sclauzero, A. P. Seitsonen, A. Smogunov, P. Umari, R. M.Wentzcovitch, J.Phys.:Condens.Matter 21 (2009) 395502-395520.
[20] D. Vanderbilt, Phys. Rev. B 41 (1990) 7892-7895.
[21] M. Born and K. Huang. Dynamical Theory of Crystal Lattices (Clarendon, Oxford, 1956).
[22] G. M. Eliashberg, Sov. Phys. JETP 11 (1960) 696-702.
[23] P. B. Allen and R. C. Dynes, Phys. Rev. B 12 (1975) 905-922.
[24] J. S. Tse, Y. Yao, and K. Tanaka, Phys. Rev. Lett. 98 (2007) 117004 [4 pages].